\documentclass[aps,prd,reprint,showpacs,amsmath,amssymb,10pt]{revtex4-1}
\usepackage{graphicx}
\usepackage{bm}
\usepackage{slashed}

\begin{document}
\title{Compton Process in Intense Short Laser Pulses}
\author{K. Krajewska}
\email[E-mail address:\;]{Katarzyna.Krajewska@fuw.edu.pl}
\author{J. Z. Kami\'nski}
\affiliation{Institute of Theoretical Physics, Faculty of Physics, University of Warsaw, Ho\.{z}a 69,
00-681 Warszawa, Poland}
\date{\today}
\begin{abstract}
The spectra of Compton radiation emitted during electron scattering off an intense laser beam 
are calculated using the framework of strong-field quantum electrodynamics. We model these intense
laser beams as finite length plane-wave-fronted pulses, similar to Neville and Rohrlich
[Phys. Rev. D {\bf 3}, 1692 (1971)], or as trains of such pulses. Expressions for energy and
angular distributions of Compton photons are derived such that a comparison of both situations becomes
meaningful. Comparing frequency distributions for both an isolated laser pulse and a laser pulse train, 
we find a very good agreement between the results for long pulse durations which breaks down however 
for ultrashort laser pulses. The dependence of angular distributions of emitted radiation on a pulse
duration is also investigated. Pronounced asymmetries of angular distributions are found for very
short laser pulses, which gradually disappear with increasing the
number of laser field oscillations. Those asymmetries are attributed to asymmetries of the
vector potential describing an incident laser beam.
\end{abstract}
\maketitle

\section{Introduction}

Compton scattering of a laser beam with a relativistic electron beam has become an efficient
source of highly polarized, intense x-ray and $\gamma$-ray radiations. These monoenergetic and
tunable Compton photon beams find numerous industrial, medical, and scientific applications. 
In particular, they bring a great deal of attention in the view of future colliders such 
as the International Linear Collider (ILC)~\cite{Araki}, the Compact Linear Collider (CLIC)~\cite{Bulyak},
and the Super B~\cite{SuperB}, which are planned to use the Compton $\gamma$ rays to produce highly polarized 
positrons. Taking into account those various applications and research interests, it becomes of great importance 
to predict theoretically spectral and spatial distributions of Compton photon beams, and to control 
their properties using the incident beam parameters.

The hypothesis of a wavelength shift of x-rays scattered by a free electron at rest
has been put forward by Compton in Ref.~\cite{Compton1}, and proved by himself experimentally~\cite{Compton2}.
Following theoretical investigations have been focused on the so-called {\it linear Compton 
scattering}~\cite{Milburn,Arutyunian,Nikishov,Narozhny,Brown,Goldman}. However, in extremely intense 
laser fields the scattering can occur in a nonlinear regime; the process known as the 
{\it nonlinear Compton scattering}. For the development of related theoretical studies, 
the reader is referred to a recent review by Ehlotzky and co-authors~\cite{Ehlotzky}. In brief, in a majority of works concerning 
the nonlinear Compton process, the incident laser beam has been described as a monochromatic plane wave field
which enables one to fully account for the electron--laser-beam interaction using 
standard methods of strong-field quantum electrodynamics (QED) (see, for instance,~\cite{Ehlotzky,Panek,Harvey,Hartin,Ivanov} 
and references quoted therein). The first analysis that goes beyond the monochromatic plane-wave approximation
when analyzing the nonlinear Compton scattering that we are aware of was by Neville and Rohrlich~\cite{Neville}.
By generalizing the usual strong-field QED methods to account for a finite length plane-wave-fronted 
pulse, the authors considered the Compton scattering by a Klein-Gordon particle. Although the method proposed
in~\cite{Neville} allows to consider QED processes in laser pulses of an arbitrary duration and strength it has not 
had a big impact until very recently. Only very few papers on the scattering of a Dirac particle
by a shaped laser pulse can be found in the literature. There is a paper by Narozhny and Fofanov~\cite{Narozhny1} who have
considered a situation when a driving laser pulse is still sufficiently long to allow to simplify
the problem significantly. In Ref.~\cite{Roshchupkin}, the same approach has been persuaded in the context 
of a resonant Compton scattering. However, for intense few-cycle laser pulses 
which are used nowadays in experimental setups, a more adequate treatment of 
the temporal structure of an incident laser pulse is necessary. This has been offered
in very recent papers~\cite{Boca1,Boca2,Seipt,Mackenroth,Heinzl,Mackenroth1} using the approach
of Neville and Rohrlich~\cite{Neville}.

In Ref.~\cite{Boca1}, Boca and Florescu have formulated the formalism for the nonlinear 
Compton scattering based on temporally shaped Volkov solutions for a Dirac particle.
The calculated spectral distributions of Compton photons were compared for different pulse 
shapes and pulse durations. In the limit of a long pulse, these results coincide with
the results of~\cite{Narozhny1} showing no significant dependence on
the precise form of the laser pulse. On the other hand, for short
laser pulses pronounced carrier-envelope phase effects in spectral distributions
of the Compton photons were demonstrated in~\cite{Boca1,Mackenroth1}. A complementary study 
for single-cycle laser pulses was presented
in~\cite{Mackenroth}, whereas the emphasis on a comparison with the {\it nonlinear
Thomson scattering} (for reviews, see, for instance~\cite{Lau,Ums}) was put in Refs.~\cite{Seipt,Boca2,Heinzl}. 
All these theoretical works follow various related experiments.
Experimentally, the nonlinear Compton scattering of relativistic electrons by 
an intense laser pulse has been observed for the first time by Englert and Rinehart~\cite{Englert}.
In another experiment, the transition between the Thomson and Compton regimes of electron scattering
has been reported~\cite{Moore}. However, the most prominent experiment showing
nonlinear effects in the Compton scattering is the SLAC experiment~\cite{Bula,Bamber}
in which the fourth harmonic of the Compton radiation was detected.

Using the approach of Neville and Rohrlich~\cite{Neville}, developed recently by
Boca and Florescu~\cite{Boca1}, we shall consider in this paper the Compton scattering of an electron
by a strong temporally shaped laser field. Our focus here is to develop new theoretical and numerical
methods to describe the situation when the incident laser field represents either an isolated laser pulse
or a train of such pulses. The sensitivity of both spectral and angular distributions of the Compton
radiation to a laser pulse duration will be studied in this paper in great detail.

Our paper is organized as follows. In Sec.~\ref{theory}, we shall introduce the theory of Compton
scattering of electrons by a temporally shaped ultrastrong laser beam. Two cases will be
considered, when the laser beam is represented by a train of laser pulses (Sec.~\ref{train})
and by an individual laser pulse (Sec.~\ref{single}). In each case, a distribution
of emitted radiation energy will be defined such that a comparison between
the two situations is possible. In Sec.~\ref{shape}, we shall define a laser pulse 
shape that will be used in our numerical calculations. In Sec.~\ref{freq}, we will
analyze frequency distributions of Compton photons for different pulse durations. The
results for an isolated laser pulse, a train of laser pulses, and for a monochromatic
laser field will be compared. Angular distributions of emitted radiation for a single
laser pulse will be shown in Sec.~\ref{angular}. The emphasis will be put there on
asymmetries in angular distributions which are observed for Compton processes induced 
by ultrashort laser pulses. Sec.~\ref{conclusions} will then be devoted to a summary of 
our results and to some final remarks.

Throughout the paper, we use the following mathematical convention and notation.
In formulas, we keep $\hbar= 1$, however our numerical results 
are presented in relativistic units such that $c = m_{\rm e} = 1$, where $m_{\rm e}$ is the electron
mass. We write $a \cdot b = a^\mu b_\mu (\mu = 0, 1, 2, 3)$ for a product of any two 
four-vectors $a$ and $b$, and $\slashed{a} = \gamma\cdot a = \gamma^\mu a_\mu$ where $\gamma^\mu$ are the 
Dirac gamma matrices. In the following, the Einstein summation convention is used.

\section{Theory}
\label{theory}

Using the $S$-matrix formalism, we find that in the lowest order of perturbation theory, the probability amplitude for the Compton process 
$e^-_{\bm{p}_{\mathrm{i}}\lambda_{\mathrm{i}}}\rightarrow e^-_{\bm{p}_{\mathrm{f}}\lambda_{\mathrm{f}}}+\gamma_{\bm{K}\sigma}$, 
with the initial and final electron momenta and spin polarizations 
$\bm{p}_{\mathrm{i}}\lambda_{\mathrm{i}}$ and $\bm{p}_{\mathrm{f}}\lambda_{\mathrm{f}}$, 
respectively, equals
\begin{equation}
{\cal A}(e^-_{\bm{p}_{\mathrm{i}}\lambda_{\mathrm{i}}}\rightarrow e^-_{\bm{p}_{\mathrm{f}}\lambda_{\mathrm{f}}}
+\gamma_{\bm{K}\sigma})=-\mathrm{i}e\int \mathrm{d}^{4}{x}\, j^{(++)}_{\bm{p}_{\mathrm{f}}\lambda_{\mathrm{f}},
\bm{p}_{\mathrm{i}}\lambda_{\mathrm{i}}}(x)\cdot A^{(-)}_{\bm{K}\sigma}(x), \label{ComptonAmplitude}
\end{equation}
where $\bm{K}\sigma$ denotes the Compton photon momentum and polarization. In the above equation,
\begin{equation}
A^{(-)}_{\bm{K}\sigma}(x)=\sqrt{\frac{1}{2\varepsilon_0\omega_{\bm{K}}V}} 
\,\varepsilon^*_{\bm{K}\sigma}\mathrm{e}^{\mathrm{i}K\cdot x},
\label{per}
\end{equation}
where $V$ is the quantization volume, $\varepsilon_0$ is the vacuum electric permittivity,
$\omega_{\bm{K}}=cK^0=c|\bm{K}|$ ($K\cdot K=0$), and $\varepsilon_{\bm{K}\sigma}=(0,\bm{\varepsilon}_{\bm{K}\sigma})$ 
is the polarization four-vector satisfying the conditions,
\begin{equation}
K\cdot\varepsilon_{\bm{K}\sigma}=0,\quad \varepsilon_{\bm{K}\sigma}\cdot\varepsilon_{\bm{K}\sigma'}=-\delta_{\sigma\sigma'},
\end{equation}
for $\sigma,\sigma'=1,2$. Moreover, $j^{(++)}_{\bm{p}_{\mathrm{f}} \lambda_{\mathrm{f}},\bm{p}_{\mathrm{i}}\lambda_{\mathrm{i}}}(x)$ 
(pluses indicate that we deal with particles of positive energy) is the matrix 
element of the electron current operator with its $\nu$-component equal to
\begin{equation}
[j^{(++)}_{\bm{p}_{\mathrm{f}} \lambda_{\mathrm{f}},\bm{p}_{\mathrm{i}}\lambda_{\mathrm{i}}}(x)]^{\nu}
=\bar{\psi}^{(+)}_{\bm{p}_{\mathrm{f}} \lambda_{\mathrm{f}}}(x)\gamma^\nu \psi^{(+)}_{\bm{p}_{\mathrm{i}}\lambda_{\mathrm{i}}}(x).
\end{equation}
Here, $\psi^{(+)}_{\bm{p}\lambda}(x)$ is the so-called Volkov solution of the Dirac equation~\cite{Volkov,KK}
\begin{equation}
\psi^{(+)}_{\bm{p}\lambda}(x)=\sqrt{\frac{m_{\mathrm{e}}c^2}{VE_{\bm{p}}}}\Bigl(1-\frac{e}{2k\cdot p}\slashed{A}\slashed{k}\Bigr)
u^{(+)}_{\bm{p}\lambda}\mathrm{e}^{-\mathrm{i} S_p^{(+)}(x)} , \label{Volk}
\end{equation}
with
\begin{equation}
S_p^{(+)}(x)=p\cdot x+\int_{-\infty}^{k\cdot x} \Bigl[\frac{ e A(\phi )\cdot p}{k\cdot p}
-\frac{e^2A^{2}(\phi )}{2k\cdot p}\Bigr]{\rm d}\phi .
\label{bbb}
\end{equation}
Moreover, $E_{\bm{p}}=cp^0$, $p=(p^0,\bm{p})$, $p\cdot p=m_{\mathrm{e}}^2c^2$, and $u^{(+)}_{\bm{p}\lambda}$ 
is the free-electron bispinor normalized such that
\begin{equation}
\bar{u}^{(+)}_{\bm{p}\lambda}u^{(+)}_{\bm{p}\lambda'}=\delta_{\lambda\lambda'}.
\end{equation}
The four-vector potential $A(k\cdot x)$ in Eq.~\eqref{Volk} represents an external electromagnetic 
radiation generated by lasers in the case when a transverse variation of the laser field 
in a focus is negligible. In this case, one can derive the exact solution to the Dirac
equation coupled to the electromagnetic field [Eq.~\eqref{Volk}], provided that 
$k\cdot A(k\cdot x)=0$ and $k\cdot k=0$.

In our further discussion, we shall adopt the Coulomb gauge for the radiation field which 
means that the four-vector $A(k\cdot x)$ has the vanishing zero component and that the 
electric and magnetic fields are equal to
\begin{align}
\bm{\mathcal{E}}(k\cdot x)&=-\partial_t \bm{A}(k\cdot x)= -ck^0 \bm{A}'(k\cdot x), \label{electric} \\
\bm{\mathcal{B}}(k\cdot x)&=\bm{\nabla}\times \bm{A}(k\cdot x)= -\bm{k}\times \bm{A}'(k\cdot x), \label{magnetic} 
\end{align}
where \textit{'prime'} means the derivative with respect to $k\cdot x$. Let us also remind that
the electric field generated by lasers has to fulfill the following condition~\cite{Becker}
\begin{equation}
\int_{-\infty}^{\infty}\bm{\mathcal{E}}(ck^0t-\bm{k}\cdot\bm{r})\mathrm{d}t=0, \label{LaserCondition1}
\end{equation}
which is followed by
\begin{equation}
\lim\limits_{t\rightarrow -\infty}\bm{A}(ck^0t-\bm{k}\cdot\bm{r})=
\lim\limits_{t\rightarrow \infty}\bm{A}(ck^0t-\bm{k}\cdot\bm{r}). \label{LaserCondition2}
\end{equation}
In the next sections, we shall consider two cases; when the vector potential $A(k\cdot x)$ describes 
a sequence of identical laser pulses (the so-called train of pulses, for which the plane 
wave is a particular realization) or a single laser pulse. Each of these situations requires a
separate theoretical treatment. 

\subsection{Train of laser pulses}
\label{train}

Let us assume that the duration of a single pulse within the field is $T_{\mathrm{p}}$. 
This means that the electromagnetic potential can be expanded as a Fourier series 
with the fundamental frequency $\omega=ck^0=2\pi/T_{\mathrm{p}}$, 
\begin{equation}
A(k\cdot x)=\sum_{N=\pm 1,\pm 2, \ldots} A_N\exp(-\mathrm{i}Nk\cdot x).
\end{equation}
Here, we do not account for the zero Fourier component since it can be eliminated by the 
gauge transformation of the vector potential. Moreover, in actual computations, we account only for a finite number 
of higher harmonics. Let us also remind that it is usually assumed that the laser field is 
adiabatically switched on in the remote past and switched off in the far future. Thus, the 
electron momentum present in Eq.~\eqref{Volk} can be interpreted as 
the field-free asymptotic momentum of the electron. Having this in mind, we consider the most 
general form of the laser field,
\begin{equation}
A(k\cdot x)=A_0\bigl[\varepsilon_1 f_1(k\cdot x)+\varepsilon_2 f_2(k\cdot x)\bigr], \label{LaserVectorPotential}
\end{equation}
in which two real four-vectors $\varepsilon_i$ describe two linear polarizations of the laser 
field such that $\varepsilon_i^2=-1$, $\varepsilon_1\cdot\varepsilon_2=0$ and $k\cdot\varepsilon_i=0$. 
Moreover, $k=k^0(1,\bm{n})$ and the normalized to 1 vector $\bm{n}$ determines the direction of 
propagation of the laser beam. Let us note that in the case of a pulse train, the so-called shape functions $f_i(\phi)$ are 
periodic functions of $\phi$ with the period equal to $2\pi$. They are however
not periodic for the case of a single laser pulse, as it will be discussed in the next Section. 
Hence, the probability amplitude for the Compton process equals
\begin{equation}
{\cal A}(e^-_{\bm{p}_\mathrm{i}\lambda_\mathrm{i}}\longrightarrow e^-_{\bm{p}_\mathrm{f}\lambda_\mathrm{f}}+\gamma_{\bm{K}\sigma})
=\mathrm{i}\sqrt{\frac{2\pi\alpha c(m_{\mathrm{e}}c^2)^2}{E_{\bm{p}_\mathrm{f}}E_{\bm{p}_\mathrm{i}}\omega_{\bm{K}}V^3}}\, \mathcal{A}, \label{ct1}
\end{equation}
where $\alpha$ is the fine-structure constant, $\alpha=e^2/(4\pi\varepsilon_0 c)$, and
\begin{align}
\mathcal{A}=& \int\mathrm{d}^{4}{x}\mathrm{e}^{-\mathrm{i}(S^{(+)}_{{p}_\mathrm{i}}(x)-S^{(+)}_{{p}_\mathrm{f}}(x)-K\cdot x)}  \label{ct2} \\
\times & \bar{u}^{(+)}_{\bm{p}_\mathrm{f}\lambda_\mathrm{f}}\Bigl(1-\mu\frac{m_\mathrm{e}c}{2p_\mathrm{f}\cdot k}\bigl[f_1(k\cdot x)\slashed{\varepsilon}_1\slashed{k}+f_2(k\cdot x)\slashed{\varepsilon}_2\slashed{k}\bigr]\Bigr) \nonumber \\
\times & \slashed{\varepsilon}^*_{\bm{K}\sigma} \Bigl(1+\mu\frac{m_\mathrm{e}c}{2p_\mathrm{i}\cdot k} \bigl[f_1(k\cdot x)\slashed{\varepsilon}_1\slashed{k}+f_2(k\cdot x)\slashed{\varepsilon}_2\slashed{k}\bigr]\Bigr) u^{(+)}_{\bm{p}_\mathrm{i}\lambda_\mathrm{i}}. \nonumber 
\end{align}
In the above equation we have introduced the important relativistically invariant parameter,
\begin{equation}
\mu=\frac{|eA_0|}{m_{\mathrm{e}}c}, \label{miu}
\end{equation}
which measures the intensity of the laser field. After some algebraic manipulations, we find that
a phase present in Eq.~\eqref{ct2} equals
\begin{equation}
S^{(+)}_{{p}_\mathrm{i}}(x)-S^{(+)}_{{p}_\mathrm{f}}(x)-K\cdot x
=(\bar{p}_\mathrm{i}-\bar{p}_\mathrm{f}-K)\cdot x+G(k\cdot x),  \label{ct3}
\end{equation}
where the so-called dressed by the laser field momentum has been introduced,
\begin{equation}
\bar{p}=p +\frac{1}{2}(\mu m_\mathrm{e} c)^2\frac{\langle f_1^2\rangle+\langle f_2^2\rangle}{p\cdot k}k . \label{ct4}
\end{equation}
We have found also that
\begin{align}
G(k\cdot x)&=\int_0^{k\cdot x}\mathrm{d}{\phi}\Bigl[ -\mu m_\mathrm{e} c\Bigl(\frac{p_\mathrm{i}\cdot\varepsilon_1}
{p_\mathrm{i}\cdot k}- \frac{p_\mathrm{f}\cdot\varepsilon_1}{p_\mathrm{f}\cdot k}\Bigr)f_1(\phi) \nonumber \\
 & -\mu m_\mathrm{e} c\Bigl(\frac{p_\mathrm{i}\cdot\varepsilon_2}{p_\mathrm{i}\cdot k}- 
\frac{p_\mathrm{f}\cdot\varepsilon_2}{p_\mathrm{f}\cdot k}\Bigr)f_2(\phi)  \nonumber \\
 &+\frac{1}{2}(\mu m_\mathrm{e} c)^2\Bigl(\frac{1}{p_\mathrm{i}\cdot k}- \frac{1}{p_\mathrm{f}\cdot k}\Bigr) \nonumber \\
 &\times (f_1^2(\phi)-\langle f_1^2\rangle + f_2^2(\phi)-\langle f_2^2\rangle) \Bigr] .  \label{ct5}
\end{align}
In the above equations, we understand that
\begin{equation}
\langle f \rangle=\frac{1}{T_{\mathrm{p}}}\int\limits_0^{T_{\mathrm{p}}}\mathrm{d}t 
f(ck^0t-\bm{k}\cdot\bm{r})=\frac{1}{2\pi}\int\limits_0^{2\pi}\mathrm{d}\phi f(\phi) . \label{ct6}
\end{equation}
Since for a train of laser pulses there is no zero Fourier component of shape functions, 
we have $\langle f_i \rangle=0$. At this point, let us emphasize that this is not in general 
true for a single laser pulse.

The advantage of applying the decomposition~\eqref{ct3} consists in the fact that $G(k\cdot x)$ 
in Eq.~\eqref{ct5} is a periodic function of its argument $k\cdot x$. Hence, we can make the following 
Fourier expansion,
\begin{equation}
[f_1(k\cdot x)]^n[f_2(k\cdot x)]^m \mathrm{e}^{-\mathrm{i}G(k\cdot x)}
=\sum_{N=-\infty}^\infty G^{(n,m)}_N \mathrm{e}^{-\mathrm{i}Nk\cdot x}. \label{ct7}
\end{equation}
For a monochromatic plane wave laser field, the Fourier coefficients $G^{(n,m)}_N$ can be expressed 
in terms of the generalized Bessel functions. For a general pulse, they are represented by a multiple sum over 
the product of many ordinary Bessel functions, and for this reason we do not present their explicit form
(see, e.g., for a bichromatic laser field in the context of laser-induced pair creation~\cite{TwoColor}).
Using the above series expansion we can now carry out the space-time integration in Eq. \eqref{ct2}.
This leads to
\begin{equation}
\mathcal{A}=\sum_N D_N \int\mathrm{d}^{4}{x} \mathrm{e}^{-\mathrm{i}(\bar{p}_\mathrm{i}+Nk-\bar{p}_\mathrm{f}-K)\cdot x}, \label{ct8}
\end{equation}
where
\begin{align}
D_N=&\bar{u}^{(+)}_{\bm{p}_\mathrm{f}\lambda_\mathrm{f}}\slashed{\varepsilon}^*_{\bm{K}\sigma} u^{(+)}_{\bm{p}_\mathrm{i}\lambda_\mathrm{i}} G^{(0,0)}_N \label{ct9}  \\
 + & \frac{1}{2}\mu m_\mathrm{e} c\Bigl[\Bigl( \frac{1}{p_\mathrm{i}\cdot k} \bar{u}^{(+)}_{\bm{p}_\mathrm{f}\lambda_\mathrm{f}}\slashed{\varepsilon}^*_{\bm{K}\sigma}  \slashed{\varepsilon}_1\slashed{k} u^{(+)}_{\bm{p}_\mathrm{i}\lambda_\mathrm{i}}
 \nonumber \\ &\qquad
 -\frac{1}{p_\mathrm{f}\cdot k} \bar{u}^{(+)}_{\bm{p}_\mathrm{f}\lambda_\mathrm{f}} \slashed{\varepsilon}_1\slashed{k}\slashed{\varepsilon}^*_{\bm{K}\sigma}  u^{(+)}_{\bm{p}_\mathrm{i}\lambda_\mathrm{i}}\Bigr)G^{(1,0)}_N  \nonumber \\
 &\qquad +\Bigl( \frac{1}{p_\mathrm{i}\cdot k} \bar{u}^{(+)}_{\bm{p}_\mathrm{f}\lambda_\mathrm{f}} \slashed{\varepsilon}^*_{\bm{K}\sigma} \slashed{\varepsilon}_2\slashed{k} u^{(+)}_{\bm{p}_\mathrm{i}\lambda_\mathrm{i}}  \nonumber \\ &\qquad
-\frac{1}{p_\mathrm{f}\cdot k} \bar{u}^{(+)}_{\bm{p}_\mathrm{f}\lambda_\mathrm{f}} \slashed{\varepsilon}_2\slashed{k}\slashed{\varepsilon}^*_{\bm{K}\sigma}  u^{(+)}_{\bm{p}_\mathrm{i}\lambda_\mathrm{i}}\Bigr)G^{(0,1)}_N\Bigr]  \nonumber \\
 - & \frac{(\mu m_\mathrm{e} c)^2}{4(p_\mathrm{i}\cdot k)(p_\mathrm{f}\cdot k)} \Bigl[\bar{u}^{(+)}_{\bm{p}_\mathrm{f}\lambda_\mathrm{f}} \slashed{\varepsilon}_1\slashed{k} \slashed{\varepsilon}^*_{\bm{K}\sigma} \slashed{\varepsilon}_1\slashed{k} u^{(+)}_{\bm{p}_\mathrm{i}\lambda_\mathrm{i}}G^{(2,0)}_N  
 \nonumber \\ &\qquad
+\bar{u}^{(+)}_{\bm{p}_\mathrm{f}\lambda_\mathrm{f}} \slashed{\varepsilon}_2\slashed{k} \slashed{\varepsilon}^*_{\bm{K}\sigma} \slashed{\varepsilon}_2\slashed{k} u^{(+)}_{\bm{p}_\mathrm{i}\lambda_\mathrm{i}}G^{(0,2)}_N  \nonumber \\
 & \qquad +\Bigl(\bar{u}^{(+)}_{\bm{p}_\mathrm{f}\lambda_\mathrm{f}} \slashed{\varepsilon}_1\slashed{k} \slashed{\varepsilon}^*_{\bm{K}\sigma} \slashed{\varepsilon}_2\slashed{k} u^{(+)}_{\bm{p}_\mathrm{i}\lambda_\mathrm{i}}
 \nonumber \\ &\qquad
  +\bar{u}^{(+)}_{\bm{p}_\mathrm{f}\lambda_\mathrm{f}} \slashed{\varepsilon}_2\slashed{k} \slashed{\varepsilon}^*_{\bm{K}\sigma} \slashed{\varepsilon}_1\slashed{k} u^{(+)}_{\bm{p}_\mathrm{i}\lambda_\mathrm{i}}  \Bigr)G^{(1,1)}_N \Bigr] .\nonumber
\end{align}
Performing now the integration in Eq. \eqref{ct8}, which leads to the four-momenta conservation condition,
\begin{equation}
\bar{p}_\mathrm{i}+Nk-\bar{p}_\mathrm{f}-K=0, \label{ct10}
\end{equation}
we arrive at the angular distribution of energy power of Compton photons that with polarization $\sigma$ 
are emitted in the space direction $\bm{n}_{\bm{K}}$, provided that the initial electron has the momentum 
$\bm{p}_{\mathrm{i}}$ and the spin polarization $\lambda_\mathrm{i}$,
\begin{align}
\frac{\mathrm{d}^{2}{P^{(\mathrm{t})}_{\mathrm{C}}(\bm{n}_{\bm{K}}\sigma; \bm{p}_\mathrm{i}\lambda_\mathrm{i})}}{\mathrm{d}{\Omega_{\bm{K}}}} =& \sum_N\sum_{\lambda_\mathrm{f}}\int \mathrm{d}^{3}{p_\mathrm{f}}\mathrm{d}{\omega_{\bm{K}}} \frac{\alpha (m_{\mathrm{e}}c^2)^2\omega^2_{\bm{K}}}{2\pi cE_{\bm{p}_{\mathrm{i}}}E_{\bm{p}_{\mathrm{f}}}} \nonumber \\ \times & |D_N|^2\delta^{(4)}(\bar{p}_\mathrm{i}+Nk-\bar{p}_\mathrm{f}-K). \label{ct11}
\end{align}
Due to the presence of the delta function, the remaining integrals can be performed exactly. Let us note first
that since $K=K^0 n_{\bm{K}}=K^0(1,\bm{n}_{\bm{K}})$, the momentum conservation condition \eqref{ct10} leads to
\begin{equation}
K^0=N\frac{k\cdot p_\mathrm{i}}{n_{\bm{K}}\cdot (\bar{p}_\mathrm{i}+Nk)}, \label{ct12}
\end{equation}
and
\begin{equation}
\bar{p}_\mathrm{f}=\bar{p}_\mathrm{i}+Nk-NK\frac{k\cdot p_\mathrm{i}}{K\cdot (\bar{p}_\mathrm{i}+Nk)}. \label{ct13}
\end{equation}
Hence, the final dressed electron momentum $\bar{p}_\mathrm{f}$ is independent of $K^0$, and
\begin{equation}
\int \mathrm{d}{K^0}\delta^{(1)}(\bar{p}_{\mathrm{i}}^0+Nk^0-\bar{p}_{\mathrm{f}}^0-K^0)=1.
\end{equation}
Moreover,
\begin{equation}
\int \mathrm{d}^3p_{\mathrm{f}}\,\delta^{(3)}(\bar{\bm{p}}_\mathrm{i}+N\bm{k}-\bar{\bm{p}}_\mathrm{f}-\bm{K})
=\Big| \frac{\partial \bm{p}_\mathrm{f}}{\partial \bar{\bm{p}}_\mathrm{f}}\Big|= \frac{p_{\mathrm{f}}^0}{\bar{p}_{\mathrm{f}}^0},
\end{equation}
and finally,
\begin{equation}
\frac{\mathrm{d}^{2}{P^{(\mathrm{t})}_{\mathrm{C}}}(\bm{n}_{\bm{K}}\sigma; \bm{p}_\mathrm{i}\lambda_\mathrm{i})}{\mathrm{d}{\Omega_{\bm{K}}}} 
= \sum_N \sum_{\lambda_\mathrm{f}} \frac{\alpha (m_{\mathrm{e}}c)^2\omega^2_{\bm{K}}}{2\pi p_{\mathrm{i}}^0 \bar{p}_{\mathrm{f}}^0} |D_N|^2 . \label{ct14}
\end{equation}
Multiplying the above equation by the duration of a single pulse $T_{\mathrm{p}}$, we obtain the energy 
distribution which is emitted in the direction $\bm{n}_{\bm{K}}$ with the polarization $\sigma$ per pulse,
\begin{equation}
\frac{\mathrm{d}^{2}{E^{(\mathrm{t})}_{\mathrm{C}}}(\bm{n}_{\bm{K}}\sigma; \bm{p}_\mathrm{i}\lambda_\mathrm{i})}{\mathrm{d}{\Omega_{\bm{K}}}} 
= T_{\mathrm{p}}\sum_N \sum_{\lambda_\mathrm{f}} \frac{\alpha (m_{\mathrm{e}}c)^2\omega^2_{\bm{K}}}{2\pi p_{\mathrm{i}}^0 \bar{p}_{\mathrm{f}}^0} |D_N|^2 . \label{ct15}
\end{equation}
For a monochromatic plane wave laser field, being a particular realization of the laser pulse
train, the integer $N$ is interpreted as a net number of laser photons 
absorbed during the process. In a more general case of a multichromatic field, the quantity $N\hbar\omega$, 
where $\omega=ck^0$, is a net energy absorbed from the laser field. This interpretation allows to define 
the angular and frequency distribution of energy emitted as Compton photons which, based on Eq.~\eqref{ct15}, is
\begin{equation}
\frac{\mathrm{d}^{2}{E^{(\mathrm{t})}_{\mathrm{C},N}}(\bm{n}_{\bm{K}}\sigma; 
\bm{p}_\mathrm{i}\lambda_\mathrm{i})}{\mathrm{d}{\Omega_{\bm{K}}}} = T_{\mathrm{p}} 
\sum_{\lambda_\mathrm{f}} \frac{\alpha (m_{\mathrm{e}}c)^2\omega^2_{\bm{K}}}{2\pi p_{\mathrm{i}}^0 \bar{p}_{\mathrm{f}}^0} |D_N|^2 . \label{ct16}
\end{equation}
If we are not interested in polarization and spin effects, then we sum up over $\sigma$ and average 
with respect to $\lambda_\mathrm{i}$,
\begin{equation}
\frac{\mathrm{d}^{2}{E^{(\mathrm{t})}_{\mathrm{C},N}}(\bm{n}_{\bm{K}}; \bm{p}_\mathrm{i})}{\mathrm{d}{\Omega_{\bm{K}}}} = \frac{1}{2}\sum_{\sigma=1,2}\sum_{\lambda_\mathrm{i}=\pm}
\frac{\mathrm{d}^{2}{E^{(\mathrm{t})}_{\mathrm{C},N}}(\bm{n}_{\bm{K}}\sigma; \bm{p}_\mathrm{i}\lambda_\mathrm{i})}{\mathrm{d}{\Omega_{\bm{K}}}}
\end{equation}
which symbolically can be written as
\begin{equation}
\frac{\mathrm{d}^{3}{E^{(\mathrm{t})}_{\mathrm{C}}}(\bm{n}_{\bm{K}}; \bm{p}_\mathrm{i})}{\mathrm{d}N\mathrm{d}{\Omega_{\bm{K}}}} =
\frac{\mathrm{d}^{2}{E^{(\mathrm{t})}_{\mathrm{C},N}}(\bm{n}_{\bm{K}}; \bm{p}_\mathrm{i})}{\mathrm{d}{\Omega_{\bm{K}}}} , \label{ComptonTrain}
\end{equation}
and interpreted as the angular and frequency (since there is one-to-one correspondence between $N$ and 
the Compton photon frequency $\omega_{\bm{K}}$) distribution of radiation energy generated during the 
process per one laser pulse.

\subsection{Single laser pulse}
\label{single}

In this Section, we will formulate theory for the Compton process by an isolated laser pulse. 
Let a pulse last for a period $T_\mathrm{p}$. This introduces the fundamental frequency 
$\omega=2\pi/T_\mathrm{p}$ and the laser field four-vector $k=k^0(1,\bm{n})$, where $\omega=ck^0$, 
with a direction of the laser pulse propagation given by the unit vector $\bm{n}$. Thus, 
the laser field potential is of the form \eqref{LaserVectorPotential}, with the same meaning 
of the symbols as above. In order to interpret the momentum $p$ in the Volkov solution 
\eqref{Volk} as the asymptotic momentum of the free electron (in both the remote past and 
the far future), we assume that
\begin{equation}
A(k\cdot x)=0 \quad \textrm{for} \quad k\cdot x < 0\quad \textrm{and}\quad k\cdot x > 2\pi. \label{LaserVectorPotentialPulse}
\end{equation}
Similar to the case of a laser pulse train, we expand the four-vector potential $A(k\cdot x)$ 
in the Fourier series. This time, however, in order to maintain 
the condition \eqref{LaserVectorPotentialPulse}, a constant term in the Fourier expansion can appear. 
Such a term will be absent in the electric field \eqref{electric}, so that the obligatory conditions, 
\eqref{LaserCondition1} and \eqref{LaserCondition2}, for the laser field can be satisfied.
At this point let us stress that the existence of the zero Fourier component in \eqref{LaserVectorPotential} 
may have profound consequences. Basically, since it leads to a non-vanishing $\langle f_i\rangle$, a
modified definition of the laser-dressed momentum must be introduced,
\begin{align}
\bar{p}=p - & \mu m_\mathrm{e} c\Bigl(\frac{p\cdot\varepsilon_1}{p\cdot k}\langle f_1\rangle 
+ \frac{p\cdot\varepsilon_2}{p\cdot k}\langle f_2\rangle\Bigr)k \nonumber \\ 
+ & \frac{1}{2}(\mu m_\mathrm{e} c)^2\frac{\langle f_1^2\rangle+\langle f_2^2\rangle}{p\cdot k}k .  \label{cp1}
\end{align}
Since now $\bar{p}$ is polarization-dependent one may expect, for instance, to observe 
asymmetries in angular distributions of the Compton photons. This will be discussed in detail
in Sec.~\ref{angular}.

Next, we have to reformulate the theory in such a way that it could effectively be used in 
numerical investigations. To this end let us go back to the space-time integral \eqref{ct2}. 
It can be expressed in term of integrals
\begin{align}
C^{(n,m)}=\int\mathrm{d}^{4}{x} & [f_1(k\cdot x)]^n[f_2(k\cdot x)]^m \label{cp2} \\  
& \times  \mathrm{e}^{-\mathrm{i}(S^{(+)}_{{p}_\mathrm{i}}(x)-S^{(+)}_{{p}_\mathrm{f}}(x)-K\cdot x)}, \nonumber
\end{align}
with $n,m=0,1,2$. By passing to the light-cone variables (Appendix \ref{LC}) we see that 
the integral over $x^-$ is limited to the finite region, $0\leqslant x^-\leqslant 2\pi/k^0$, 
provided that $n$ and $m$ are not simultaneously equal to 0. Hence, in order to determine numerically 
$C^{(0,0)}$ we have to transform this integral to a more suitable form; this is done by applying the 
Boca-Florescu transformation presented in the Appendix \ref{BF}. We have the following correspondences:
\begin{equation}
Q=p_{\mathrm{i}}-p_{\mathrm{f}}-K ,
\end{equation}
and
\begin{equation}
h(\phi)=a_1 f_1(\phi)+ a_2 f_2(\phi)+b[f^2_1(\phi)+f^2_2(\phi)] ,
\end{equation}
with
\begin{align}
a_1=& -\mu m_{\mathrm{e}}c\Bigl(\frac{p_{\mathrm{i}}\cdot\varepsilon_1}{p_{\mathrm{i}}\cdot k} - \frac{p_{\mathrm{f}}\cdot\varepsilon_1}{p_{\mathrm{f}}\cdot k}\Bigr)=-Q^0\tilde{a}_1/k^0 , \nonumber \\
a_2=& -\mu m_{\mathrm{e}}c\Bigl(\frac{p_{\mathrm{i}}\cdot\varepsilon_2}{p_{\mathrm{i}}\cdot k} - \frac{p_{\mathrm{f}}\cdot\varepsilon_2}{p_{\mathrm{f}}\cdot k}\Bigr)=-Q^0\tilde{a}_2/k^0 , \nonumber \\
b=& \frac{1}{2}(\mu m_{\mathrm{e}}c)^2 \Bigl(\frac{1}{p_{\mathrm{i}}\cdot k} - \frac{1}{p_{\mathrm{f}}\cdot k}\Bigr)=-Q^0\tilde{b}/k^0 .
\end{align}
which also defines parameters $\tilde{a}_i$ and $\tilde{b}$, provided that $Q^0\neq 0$. 
We shall demonstrate below that this condition is always fulfilled [see, Eq. \eqref{cp9} 
below]. Finally, applying Eq. \eqref{BocaFlorescu} we arrive at
\begin{align}
C^{(0,0)}= &\int\mathrm{d}^{4}{x}
\Bigl(\tilde{a}_1 f_1(k\cdot x)+ \tilde{a}_2 f_2(k\cdot x)+\tilde{b}[f^2_1(k\cdot x)
\nonumber \\
 &\qquad +f^2_2(k\cdot x)]\Bigr) \mathrm{e}^{-\mathrm{i}(S^{(+)}_{{p}_\mathrm{i}}(x)-S^{(+)}_{{p}_\mathrm{f}}(x)-K\cdot x)} .
\label{aaa}
\end{align}
The next steps basically follow the procedure of Sec.~\ref{train}. 
We make the decomposition, similar to Eq. \eqref{ct3},
\begin{equation}
S^{(+)}_{{p}_\mathrm{i}}(x)-S^{(+)}_{{p}_\mathrm{f}}(x)-K\cdot x=(\bar{p}_\mathrm{i}-\bar{p}_\mathrm{f}-K)\cdot x+G(k\cdot x),  \label{cp3}
\end{equation}
with the dressed momenta defined by Eq. \eqref{cp1}, and
\begin{align}
G(k\cdot x)&=\int_0^{k\cdot x}\mathrm{d}{\phi}\Bigl[ -\mu m_\mathrm{e} c\Bigl(\frac{p_\mathrm{i}\cdot\varepsilon_1}{p_\mathrm{i}\cdot k}- \frac{p_\mathrm{f}\cdot\varepsilon_1}{p_\mathrm{f}\cdot k}\Bigr)\nonumber\\
&\times \bigl(f_1(\phi) -\langle f_1\rangle\bigr) -\mu m_\mathrm{e} c\Bigl(\frac{p_\mathrm{i}\cdot\varepsilon_2}{p_\mathrm{i}\cdot k}- \frac{p_\mathrm{f}\cdot\varepsilon_2}{p_\mathrm{f}\cdot k}\Bigr)\nonumber\\
&\times\bigl(f_2(\phi)-\langle f_2\rangle\bigr) 
 +\frac{1}{2}(\mu m_\mathrm{e} c)^2\Bigl(\frac{1}{p_\mathrm{i}\cdot k}- \frac{1}{p_\mathrm{f}\cdot k}\Bigr)\nonumber\\
&\times\bigl(f_1^2(\phi)-\langle f_1^2\rangle + f_2^2(\phi)-\langle f_2^2\rangle\bigr) \Bigr] .  \label{cp4}
\end{align}
Now, after applying the Fourier decompositions \eqref{ct7}, we obtain $D_N$ similar to \eqref{ct9} with the only replacement
\begin{equation}
G^{(0,0)}_N \rightarrow \tilde{a}_1 G^{(1,0)}_N+ \tilde{a}_2 G^{(0,1)}_N+\tilde{b}[G^{(2,0)}_N+G^{(0,2)}_N] ,
\end{equation}
which follows from the application of the Boca-Florescu transformation with respect to Eq.~\eqref{aaa}.

Performing the space-time integration in Eq. \eqref{ct2} and keeping in mind 
that $0\leqslant x^- \leqslant 2\pi/k^0$, we arrive at
\begin{equation}
\mathcal{A}=\sum_N (2\pi)^3\delta^{(1)}(P_N^-)\delta^{(2)}(\bm{P}_N^\bot)D_N\frac{1-\mathrm{e}^{-2\pi\mathrm{i}P_N^+/k^0}}{\mathrm{i}P_N^+} ,  \label{cp5}
\end{equation}
where
\begin{equation}
P_N=\bar{p}_{\mathrm{i}}+Nk-\bar{p}_{\mathrm{f}}-K . \label{cp6}
\end{equation}
In order to solve the momentum conservation conditions imposed by the three delta functions, let us introduce the four-vector $w=p_{\mathrm{i}}-K$, so that
\begin{equation}
p_{\mathrm{f}}^0=p_{\mathrm{f}}^{\|}+w^-, \quad \bm{p}_{\mathrm{f}}^{\bot}=\bm{w}^{\bot}. \label{cp7}
\end{equation}
Since the electron mass is different from zero, it follows from the first equation that $w^->0$, and
\begin{equation}
p_{\mathrm{f}}^{\|}=\frac{(m_{\mathrm{e}}c)^2-(w^-)^2+\bm{w}^2_{\bot}}{2w^-} =\frac{K\cdot p_{\mathrm{i}}}{w^-}+w^{\|} , \label{cp8}
\end{equation}
which means that
\begin{equation}
Q^0=p^0_{\mathrm{i}}-p^0_{\mathrm{f}}-K^0=p_{\mathrm{i}}^{\|}-p_{\mathrm{f}}^{\|}-K^{\|}=-\frac{p_{\mathrm{i}}\cdot K}{w^-}<0 . \label{cp9}
\end{equation}
This is exactly the applicability condition for the Boca-Florescu transformation \eqref{BocaFlorescu}.

Since $\bm{P}_N^\bot$ and $P_N^{-}$ do not depend explicitly on $N$, and
\begin{equation}
\int\mathrm{d}^{3}{p_{\mathrm{f}}}\,\delta^{(1)}(P_N^{-})\delta^{(2)}(\bm{P}_N^\bot)=\frac{k^0p_{\mathrm{f}}^0}{k\cdot p_{\mathrm{f}}} ,
\end{equation}
we obtain the differential distribution of energy
\begin{align}
\frac{\mathrm{d}^{3}{E^{(\mathrm{p})}_{\mathrm{C}}(\bm{K}\sigma; \bm{p}_\mathrm{i}\lambda_\mathrm{i})}}{\mathrm{d}{\omega_{\bm{K}}}
\mathrm{d}^{2}{\Omega_{\bm{K}}}} = &\sum_{\lambda_{\mathrm{f}}}\frac{\alpha (m_{\mathrm{e}}c)^2 k^0 (K^0)^2}{(2\pi)^2 p_{\mathrm{i}}^0
(k\cdot p_{\mathrm{f}})} \label{cp10} \\  \times & \Big|\sum_ND_N\frac{1-\mathrm{e}^{-2\pi\mathrm{i}P_N^0/k^0}}{P_N^0} \Big|^2 , \nonumber
\end{align}
emitted as Compton photons by a single pulse, or if we are not interested in polarization effects,
\begin{equation}
\frac{\mathrm{d}^{3}{E^{(\mathrm{p})}_{\mathrm{C}}(\bm{K}; \bm{p}_\mathrm{i})}}{\mathrm{d}{\omega_{\bm{K}}}\mathrm{d}^{2}{\Omega_{\bm{K}}}} = \frac{1}{2}\sum_{\sigma=1,2}\sum_{\lambda_\mathrm{i}=\pm}\frac{\mathrm{d}^{3}{E^{(\mathrm{p})}_{\mathrm{C}}(\bm{K}\sigma; \bm{p}_\mathrm{i}\lambda_\mathrm{i})}}{\mathrm{d}{\omega_{\bm{K}}}\mathrm{d}^{2}{\Omega_{\bm{K}}}} . \label{cp11}
\end{equation}

Here, the question arises: How to define the energy distribution of Compton photons 
scattered by a single laser pulse so it is meaningful to compare it with the 
distribution \eqref{ComptonTrain}? To answer this question let us go back to Eq. \eqref{cp5} 
and note that for a very long pulse the maximum of $|\mathcal{A}|^2$ is achieved for $P_N^+=P_N^0=0$. 
This allows us to define the effective energy $N_{\mathrm{eff}}\hbar ck^0$ absorbed from a laser 
pulse as compared to the energy $N\hbar ck^0$ absorbed from a train of laser pulses such 
that $P_{N_{\mathrm{eff}}}^0=0$, i.e.,
\begin{equation}
N_{\mathrm{eff}}=\frac{K^0+\bar{p}_{\mathrm{f}}^0-\bar{p}_{\mathrm{i}}^0}{k^0} =cT_{\mathrm{p}}\frac{K^0+\bar{p}_{\mathrm{f}}^0-\bar{p}_{\mathrm{i}}^0}{2\pi} . \label{cp12}
\end{equation}
Hence, we can define an analogue of \eqref{ComptonTrain} for an isolated laser pulse,
\begin{equation}
\frac{\mathrm{d}^{3}{E^{(\mathrm{p})}_{\mathrm{C}}(\bm{K}; \bm{p}_\mathrm{i})}}{\mathrm{d}{N_{\mathrm{eff}}}\mathrm{d}^{2}{\Omega_{\bm{K}}}}=
\frac{\mathrm{d}\omega_{\bm{K}}}{\mathrm{d}N_\mathrm{eff}} \frac{\mathrm{d}^{3}{E^{(\mathrm{p})}_{\mathrm{C}}(\bm{K}; \bm{p}_\mathrm{i})}}{\mathrm{d}{\omega_{\bm{K}}}\mathrm{d}^{2}{\Omega_{\bm{K}}}} . 
\label{ComptonPulse} 
\end{equation}
We do not present here the explicit form for the derivative ${\mathrm{d}\omega_{\bm{K}}}/{\mathrm{d}N_\mathrm{eff}}$
which is rather lengthy and can be determined numerically in a more efficient way. The aim of this paper, 
among others, is to discuss for how long laser pulses two formulas given by Eqs. \eqref{ComptonTrain} 
and \eqref{ComptonPulse} provide similar results.

\section{Shape functions}
\label{shape}

\begin{figure}
\includegraphics[width=6.5cm]{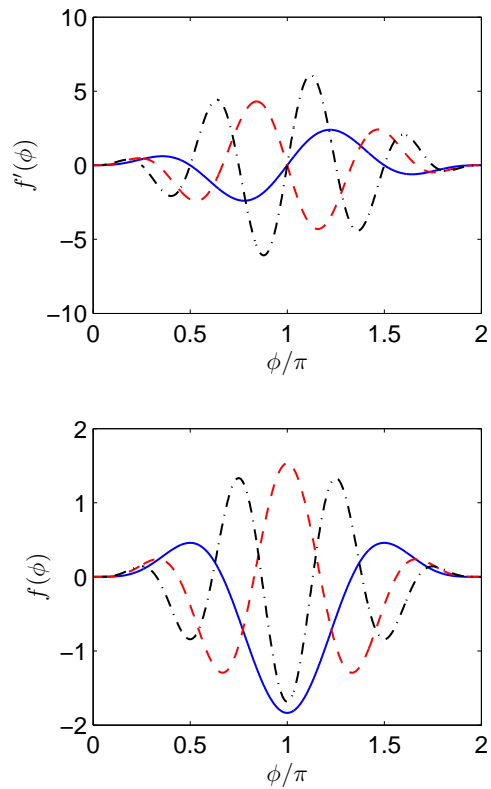}%
\caption{(Color online) Laser field shape functions [defined by Eqs. \eqref{shapefunctionE} 
and \eqref{shapefunctionA}] 
normalized according to Eq. \eqref{normalizacja}. Each curve relates to a different
number of laser field oscillations, namely, $N_{\mathrm{osc}}=2$ (solid blue line), 
$N_{\mathrm{osc}}=3$ (dash-dashed red line), and $N_{\mathrm{osc}}=4$ 
(dash-dotted black line). \label{shapes}}
\end{figure}

In our further investigations, we shall consider linearly polarized laser pulses which propagate in the 
$z$-direction ($\bm{n}=\bm{e}_z$), with a polarization along the $x$-axis 
($\bm{\varepsilon}_1=\bm{e}_x$). This means that $f_2(\phi)=0$, and so we shall denote the 
first shape function $f_1(\phi)$ simply as $f(\phi)$. 
This function has to be chosen such that the electric field satisfies the condition 
\eqref{LaserCondition1}. Moreover, we shall assume that the laser pulse lasts for a finite 
time $T_\mathrm{p}$. This excludes commonly used shape functions with envelops proportional 
to functions gradually decreasing in the remote past and the far future, like for instance 
the Gaussian function or some rational combinations of the hyperbolic functions~\cite{Boca1,Boca2,Seipt,Heinzl,Mackenroth1} (we also 
exclude rectangular pulses as unphysical). This will allow us to define without an ambiguity 
the number of oscillations within the pulse (as opposed, for instance, to Ref.~\cite{Boca1}). There are many possibilities for choosing such 
functions, but for the purpose of this publication we consider the following one-parameter shape function, 
\begin{align}
f^{\prime}(k\cdot x)= & N_{\mathrm{A}}\sin^2\Bigl(\frac{k_{\mathrm{L}}\cdot x}{2N_{\mathrm{osc}}}\Bigr)\sin(k_{\mathrm{L}}\cdot x) \nonumber \\
 = & N_{\mathrm{A}}\sin^2\Bigl(\frac{k\cdot x}{2}\Bigr)\sin(N_{\mathrm{osc}}k\cdot x), \label{shapefunctionE}
\end{align}
for $0\leqslant k_{\mathrm{L}}\cdot x\leqslant 2\pi N_{\mathrm{osc}}$, and 0 otherwise. 
The pulse duration equals $T_{\mathrm{p}}=2\pi N_{\mathrm{osc}}/\omega_{\mathrm{L}}$, hence 
$k_{\mathrm{L}}=(\omega_{\mathrm{L}}/c)(1,\bm{n})=N_{\mathrm{osc}}k$. Moreover, $\omega_{\mathrm{L}}$ 
is the central frequency of the laser pulse.

Let us mention that the shape function \eqref{shapefunctionE} determines both the electric and magnetic fields 
of the laser pulse, Eqs. \eqref{electric} and \eqref{magnetic}, respectively. Hence, the shape function 
for the four-vector potential equals
\begin{equation}
f(k\cdot x)=\int_0^{k\cdot x} \mathrm{d}\phi f^{\prime}(\phi), \label{shapefunctionA}
\end{equation}
and vanishes for $k\cdot x < 0$ and $k\cdot x > 2\pi$. In Eq.~\eqref{shapefunctionE},
the free parameter, $N_{\mathrm{osc}}=1,2,\ldots$, determines the number of oscillations 
within the pulse. The normalization constant $N_{\mathrm{A}}$ is defined such that
\begin{equation}
\langle (f-f_0)^2 \rangle=\frac{1}{2\pi}\int_0^{2\pi} (f(\phi)-f_0)^2 \mathrm{d}\phi=\frac{1}{2}, \label{normalizacja}
\end{equation}
in order to establish the connection with the monochromatic plane wave approximation. 
In the above equation, $f_0$ is the constant term in the Fourier expansion of $f(\phi)$. 
Both shape functions, $f'(k\cdot x)$ and $f(k\cdot x)$, for some small $N_{\mathrm{osc}}$ 
are presented in Fig. \ref{shapes}. As one can anticipate from this figure, for the 
given values of $N_{\rm osc}$, $\langle f\rangle\neq 0$. However, with increasing the
number of laser field oscillations $\langle f\rangle$ starts to deteriorate, and finally
goes to 0. Therefore, we expect that for sufficiently long laser pulses the energy spectrum of
Compton photons will approach the one for a train of laser pulses. This will be illustrated 
in the next Section.

We want to emphasize that the aforementioned discussion relates to a single
laser pulse. For a train of pulses, the only difference is that the shape function for 
the electric and magnetic fields, given by Eq.~\eqref{shapefunctionE}, is repeated
for all times. This means that for the pulse train, $f'(k\cdot x)$ and $f(k\cdot x)$ are
periodic functions of their argument, with vanishing zero Fourier components.

\section{Frequency distributions}
\label{freq}

\begin{figure}
\includegraphics[width=6.5cm]{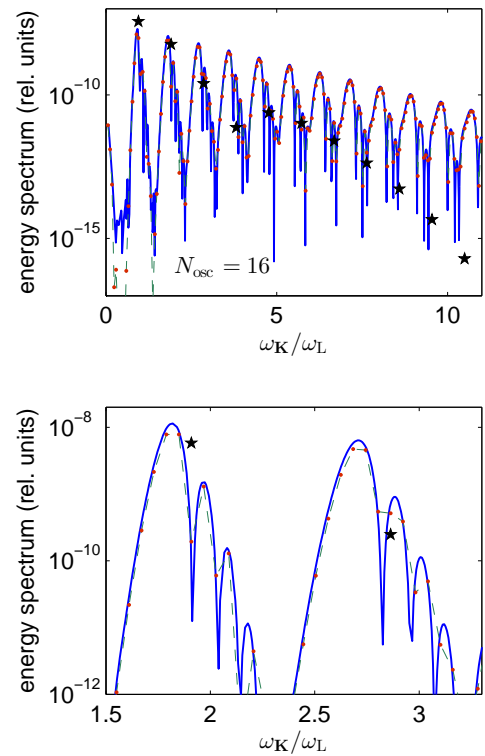}%
\caption{(Color online) Energy spectra, Eqs. \eqref{ComptonTrain} and \eqref{ComptonPulse}, 
for Compton photons emitted in the direction $\bm{n}_{\bm{K}}$ determined by the polar and 
azimuth angles $\theta_{\bm{K}}=0.2$ and $\varphi_{\bm{K}}=0$ (in radians). The initial electron 
is at rest and the laser photon frequency is such that $\hbar\omega_\mathrm{L}=3\times 10^{-6}m_\mathrm{e}c^2$.
The intensity of the laser beam is determined by $\mu=1$. The solid (blue) line corresponds 
to scattering by a single laser pulse, Eq. \eqref{ComptonPulse}, the dashed (green) line to 
scattering by a train of pulses, 
Eq. \eqref{ComptonTrain}, whereas the red bullets indicate the Compton photon frequencies, 
$\omega_{\bm{K}}=cK^0$ in Eq. \eqref{ct12}, for integer $N$. The black pentagrams represent 
the energy spectrum \eqref{ComptonTrain} for a plane-wave, when the shape function in Eq. 
\eqref{shapefunctionE} is proportional to $\sin(k_{\mathrm{L}}\cdot x)$. In all cases $N_{\mathrm{osc}}=16$. 
The lower panel shows the enlarged part of the upper one in order to prove a very good agreement 
between the results obtained for a single pulse and a train of such pulses.  \label{q1c01d2012.02.25}}
\end{figure}

\begin{figure}
\includegraphics[width=6.5cm]{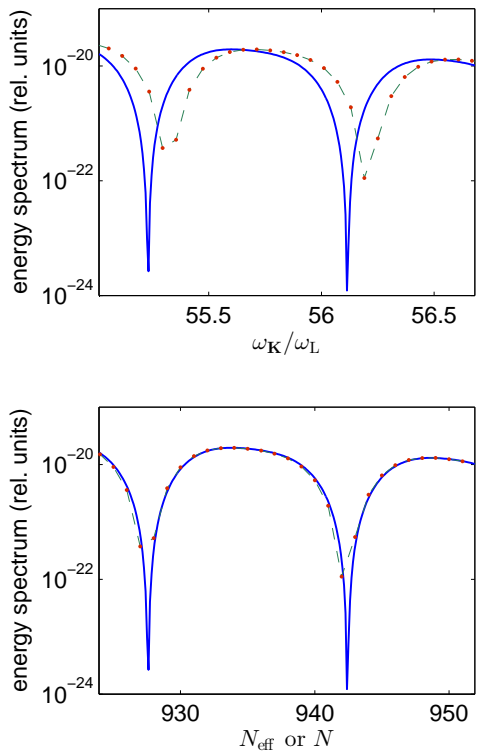}%
\caption{(Color online) The same as in Fig. \ref{q1c01d2012.02.25} but for larger frequencies
(upper panel). In the upper panel, we clearly see the 'blue-shift' of the spectrum 
for Compton photons scattered by a train of pulses, as compared to those scattered
by an individual pulse. This shift is absent, however, when
the spectra are shown as functions of $N_{\mathrm{eff}}$ or $N$ (lower panel).  
\label{q1c01ad2012.02.25}}
\end{figure}

\begin{figure}
\includegraphics[width=6.5cm]{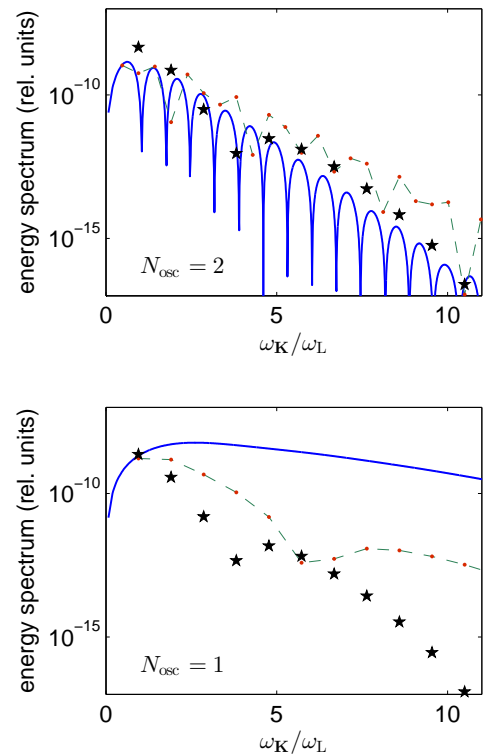}%
\caption{ (Color online) The same as in Fig. \ref{q1c01d2012.02.25} but for very short pulses with $N_{\mathrm{osc}}=2$ 
(upper panel) and $N_{\mathrm{osc}}=1$ (lower panel). Let us note that pentagrams, representing the 
results for the monochromatic plane wave, in both these panels and in Fig. \ref{q1c01d2012.02.25} 
provide the same results up to multiplication by the number of oscillations $N_\mathrm{osc}$. 
\label{q23c01d2012.02.25}}
\end{figure}

We start with discussing the frequency distribution of Compton radiation for a given 
$\bm{n}_{\bm{K}}$ and for the case when electrons are initially at rest. For the laser field we choose the one 
generated by the Ti:Sapphire laser with central frequency $\omega_{\mathrm{L}}$ such that 
$\hbar\omega_{\mathrm{L}}=1.5\textrm{eV}\approx 3\times 10^{-6}m_\mathrm{e}c^2$ (this corresponds 
approximately to wavelength 800nm). Such a choice is motivated by the fact that presently
these are the most powerful lasers generating electromagnetic radiation in the visible part 
of the spectrum. For very long laser pulses, the parameter $\mu$ introduced in \eqref{miu} 
equals 1 for the laser field intensity of the order of $10^{18}\,\textrm{W/cm}^2$. It is instructive to realize
that nowadays intensities of the order of $(10^{20}\div 10^{22})\,\textrm{W/cm}^2$ are available experimentally
\cite{Mourou}, and those correspond roughly to $\mu\approx 7\div 70$.

In Fig. \ref{q1c01d2012.02.25}, we compare energy distributions of Compton photons scattered either
by a single laser pulse, Eq. \eqref{ComptonPulse}, by a train of such pulses, Eq. \eqref{ComptonTrain}, 
and by a plane laser wave (which, in fact, is a train of 
pulses with a constant envelope) for the case when $\mu=1$ and $N_\mathrm{osc}=16$. We observe 
here a perfect agreement of the first two approaches and a failure of the monochromatic 
wave approximation. This agreement is also attained for much larger Compton frequencies provided 
that the spectra are plotted as functions of energy absorbed from the laser field (i.e., 
$N_\mathrm{eff}$ for an isolated laser pulse and $N$ for a train of pulses); see, the lower panel of 
Fig.~\ref{q1c01ad2012.02.25}. On contrary, when the frequency dependence of emitted photons
is investigated, a shift of energy spectra is observed (see, the upper panel of Fig. \ref{q1c01ad2012.02.25}). 
The shift is caused by a different dressing of the initial and final electron momenta for these two cases, 
which we have discussed above. As expected, for longer pulses this discrepancy gradually disappears 
since the constant term in the Fourier expansion of the electromagnetic potential for a single pulse 
becomes gradually smaller; this is confirmed by our results for even larger $N_{\rm osc}$
which we, however, do not present here. Instead, in Fig. \ref{q23c01d2012.02.25} we show frequency 
distributions of emitted photons for extremely short laser pulses such that there is either
two or even one oscillation within the pulse ($N_{\rm osc}=2$ or 1, respectively). 
In both these cases, a significant disagreement of frequency distributions of Compton radiation
for an individual pulse and a laser pulse train is observed, in addition to a
pronounced discrepancy with the results obtained using the monochromatic plane-wave approximation. We 
confirm therefore that for ultrashort laser pulses available these days, a precise theoretical treatment 
of their temporal properties is indeed necessary.

It is worth noting the existence of regular sidelobes in energy spectra
of Compton photons, as shown in Fig.~\ref{q1c01d2012.02.25}. 
Namely, apart from a dominating peak in the spectrum there are also
subpeaks located asymmetrically on its right-hand side, i.e., for larger
frequencies. The exactly same structure has been already observed and interpreted 
as the interference phenomenon in Ref.~\cite{Heinzl} (see, also references therein).

In closing this Section, let us comment on the momentum dressing in strong-field QED processes.
In the case of a monochromatic laser field, such a dressing is gained when separating the
classical action $S_p^{(+)}(x)$ [Eq.~\eqref{bbb}] into an oscillatory and a linear parts. A similar approach
has been proposed here, even though for finite laser pulses the respective periodicity of a
four-vector potential is lost. The reason that we have introduced it anyway is that the
concept of the momentum dressing allows 
to define a convenient method for performing numerical calculations. In addition, a perfect 
agreement between the results for a single laser pulse and a laser pulse train observed
in the lower panel of Fig.~\ref{q1c01ad2012.02.25} indicates, that the concept of quasi-momenta 
is still valid for a 16-cycle laser pulse such that $\mu=1$. As we have checked this for 
the same pulse duration but for stronger laser fields (larger $\mu$), this agreement starts 
to deteriorate and the concept of momenta dressing breaks down eventually. This may have 
profound consequences, particularly when analyzing the laser-induced pair creation process,
as it has been pointed out in Ref.~\cite{Heinzl1}.

\section{Angular distributions}
\label{angular}

\begin{figure}
\includegraphics[width=6.5cm]{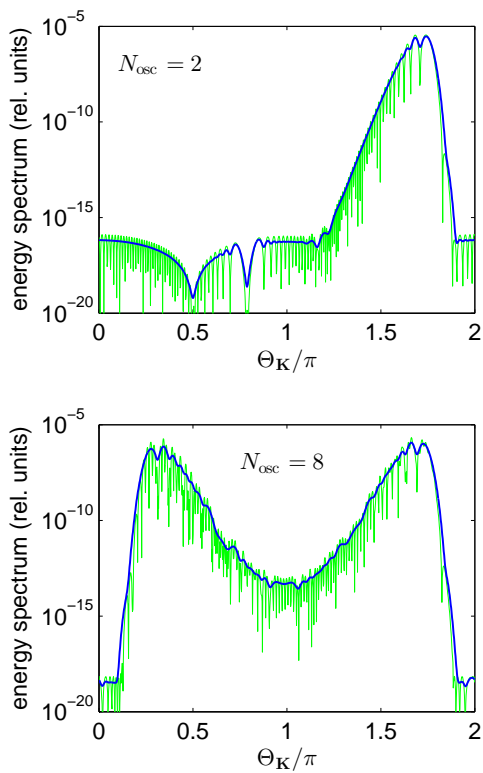}%
\caption{(Color online) Angular distributions of the Compton radiation energy (green lines), 
Eq. \eqref{ComptonPulse}, as functions of the angle $\Theta_{\bm{K}}$ in the plane spanned by the laser field
propagation direction and the polarization vector (i.e., in the $xz$-plane) for an ultrashort ($N_{\mathrm{osc}}=2$, 
upper panel) and a relatively long ($N_{\mathrm{osc}}=8$, lower panel) laser pulses. The laser field 
central frequency $\omega_\mathrm{L}$ is such that $\hbar\omega_\mathrm{L}=3\times 10^{-4}m_{\mathrm{e}}c^2$ 
whereas $\mu=1$. The initial electron is at rest and the Compton photon energy equals $\hbar\omega_{\bm{K}}= 10^{-2}m_{\mathrm{e}}c^2$. 
The smooth blue lines represent the averaged distributions, as described by Eq. \eqref{averagedangular}. 
\label{smooth2and8d2012.02.03}}
\end{figure}

\begin{figure}
\includegraphics[width=8cm]{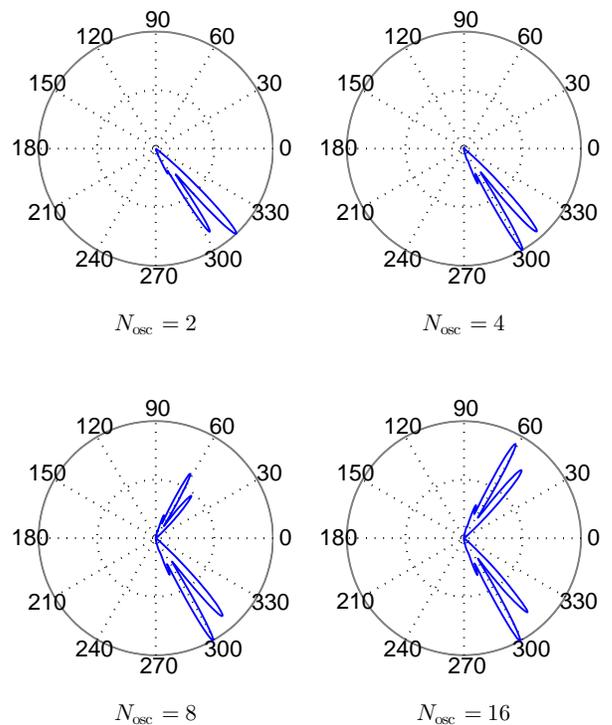}%
\caption{(Color online) Polar plots of the (normalized to 1 and window-averaged) Compton photon energy distribution, 
Eq. \eqref{averagedangular}, as functions of $\Theta_{\bm{K}}$ for different-cycle laser pulses with 
$N_{\mathrm{osc}}=2,4,8$, and 16, as indicated under the plots. The other parameters are exactly the same as in Fig.
\ref{smooth2and8d2012.02.03}.
\label{smoothpolarall2012.02.03}}
\end{figure}

When discussing the angular distributions of Compton photons, it is more convenient to introduce a polar angle $\Theta_{\bm{K}}$ 
such that it is equal to $\theta_{\bm{K}}$ for $0\leqslant \varphi_{\bm{K}}\leqslant\pi$, 
and $\Theta_{\bm{K}}=2\pi-\theta_{\bm{K}}$ for the azimuthal angle $\varphi_{\bm{K}}+\pi$. 
With this definition, the distribution \eqref{ComptonPulse} is a periodic function of $\Theta_{\bm{K}}$.
Usually, the angular distribution of the Compton radiation is a very rapidly changing function of 
$\Theta_{\bm{K}}$ for a given $\varphi_{\bm{K}}$. These changes cannot be observed experimentally because of the 
finite angular resolution of detectors. In order to account for this fact we have to average the 
spectrum \eqref{ComptonPulse} with some function, the so-called window function, concentrated around 
a given value of the polar angle $\Theta_{\bm{K}}$. Let us choose the window function as
\begin{equation}
W_{\beta}(\theta)=N_{\mathrm{W}}\Bigl(\frac{1+\cos\theta}{2} \Bigr)^{\beta} , \label{window}
\end{equation}
with $\beta>0$, which is normalized such that
\begin{equation}
\int_{0}^{2\pi}\mathrm{d}\theta W_{\beta}(\theta)=1.
\end{equation}
The parameter $\beta$ defines the half-width of this function equal to the angular resolution of the detector. 
Let us assume that the angular resolution of a detector is roughly $0.02\pi$ radians, which is achieved if $\beta=2048$. 
Then, the average of a periodic function $g(\theta)$ with the period $2\pi$ is defined as
\begin{equation}
\langle g\rangle_{W_{\beta}}(\Theta_{\bm{K}})=\int_{0}^{2\pi}\mathrm{d}\theta W_{\beta}(\Theta_{\bm{K}}-\theta)g(\theta). \label{averagedangular}
\end{equation}
This integration can be efficiently carried out by applying the Fast Fourier Transform. 
In our analysis, $g(\theta)$ is the differential energy distribution \eqref{ComptonPulse} for a given 
$\omega_{\bm{K}}$ and for a given azimuthal angle $\varphi_{\bm{K}}$.

In Fig.~\ref{smooth2and8d2012.02.03}, we show angular distributions of the Compton radiation
energy (green lines) which is emitted during electron scattering by a single laser pulse. While the upper
panel shows the results for an ultrashort laser pulse with $N_{\mathrm{osc}}=2$, the lower panel is
for a relatively long laser pulse such that $N_{\mathrm{osc}}=8$. In both cases, the central
laser frequency is such that $\hbar\omega_{\mathrm{L}}=3\times 10^{-4}m_{\mathrm{e}}c^2$,
and $\mu$ equals 1. As before, we assume that initially the colliding electron is at rest.
The presented distributions are for a fixed energy of Compton photons such that 
$\hbar\omega_{\bm{K}}= 10^{-2}m_{\mathrm{e}}c^2$ and for $\varphi_{\bm{K}}=0$. As was anticipated above,
we observe very rapid oscillations of angular distributions, and so we plot
also their average \eqref{averagedangular} which is represented 
by a blue line. For a two-cycle laser pulse, a strong asymmetry in angular distributions 
 \eqref{ComptonPulse} and \eqref{averagedangular} is observed, which is the consequence of asymmetry 
of the shape function $f(\phi)$ with respect to the change of the polarization vector direction 
$\bm{\varepsilon}_1\rightarrow -\bm{\varepsilon}_1$. Let us note that in the current case, the shape function of the electric field 
$f'(\phi)$ does not possess such an asymmetry. This indicates that symmetry properties of the electromagnetic 
potential ({\it not} the electric field) are responsible for symmetries of energy angular distributions of Compton photons. 
In Fig.~\ref{smoothpolarall2012.02.03}, we present only window-averaged angular distributions 
\eqref{averagedangular} for laser pulses of different duration, namely, $N_{\mathrm{osc}}=2,4,8$, and 16,
and for the same parameters as in Fig.~\ref{smooth2and8d2012.02.03}. Once again, we see that
the angular asymmetry of those distributions, dominant for very short laser pulses, gradually disappears
with increasing the number of laser field oscillations $N_{\rm osc}$. For $N_{\rm osc}=16$, 
the angular distribution is roughly symmetric.
We would like also to note that the corresponding distributions for Compton scattering by positrons 
are similar to Figs.~\ref{smooth2and8d2012.02.03} and~\ref{smoothpolarall2012.02.03}, however they are 
mirrored with respect to $\Theta_{\bm{K}}=\pi$.

\begin{figure}
\includegraphics[width=8.7cm]{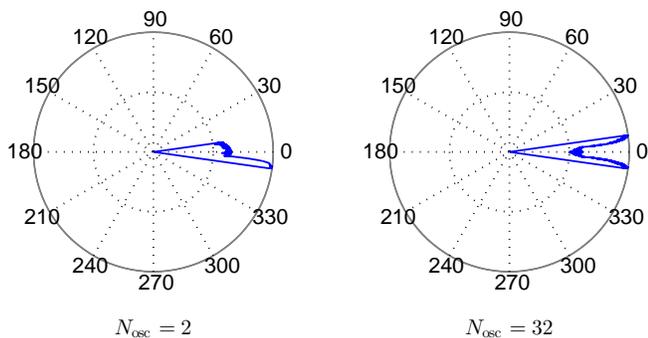}
\caption{(Color online) Polar plots of the normalized to one Compton radiation energy 
distribution~\eqref{ComptonPulse} ($\hbar\omega_{\bm K}=100 m_{\rm e}c^2)$ as a function of $\Theta_{\bm{K}}$ 
for 2- and 32-cycle laser pulses (left and right panels, respectively). Here,
$\hbar\omega_{\rm L}=m_{\rm e}c^2$ (in the electron rest frame) and $\mu=10$. For visual purposes, 
the rates have been raised to power 0.01. \label{smooth2and32mod2a}}
\end{figure}

Let us emphasize that all the results presented here relate to the case when 
an initial electron is at rest. In experiments, however, we deal with a situation when 
incident electrons move with very high kinetic energies; in the SLAC experiment reaching almost 
50 GeV that corresponds to the Lorentz factor $\gamma$ of the order of $10^5$~\cite{Bamber,Bula}. 
In Refs.~\cite{Bamber,Bula}, these energetic electrons were colliding 
with a counterpropagating laser beam to produce highly energetic radiation. 
We would like to emphasize that the calculations performed in this paper 
remain valid in the rest frame of the electron from the countermoving beam, if we
scale the fundamental frequency of the laser field by roughly a factor of $2\gamma$.
This means that for a relativistic electron that moves with a large Lorentz 
factor $\gamma$ and collides, for instance, with a Ti:Sapphire laser field, the Doppler upshifted 
frequency of the laser field in the electron rest frame can reach $m_{\rm e}c^2$.
For this reason, in Fig.~\ref{smooth2and32mod2a} we present angular distributions of
the Compton radiation for the case when in the electron rest frame the laser field central
frequency is such that $\hbar\omega_{\rm L}=m_{\rm e}c^2$, and also $\mu=10$. These distributions
are for the Compton photons of energy $\hbar\omega_{\bm K}=100 m_{\rm e}c^2$ (in the chosen
reference frame) and they have been normalized to 1. Let us note that the probability
rates of pair creation in the forward direction are significantly smaller (by roughly
twenty orders of magnitude) than the offside lobes in angular distributions. For visual
purposes, however, we have raised the data to power 0.01 so even small details of the
spectra become visible. While in the right panel of Fig.~\ref{smooth2and32mod2a} the results
for a 32-cycle laser pulse are presented, in the left panel we show the results
for a much shorter pulse which contains only two oscillations of the laser field.
Even for much stronger fields than considered previously, we observe a very clear asymmetry in 
angular distributions of Compton photons for very short laser pulses. As before, this asymmetry 
is absent for long pulse durations. We would also like to point out that 
the angular distributions shown in Fig.~\ref{smooth2and32mod2a} have not been averaged 
according to the prescription defined by Eq.~\eqref{averagedangular}. As one can observe for the given 
parameters, there is a small angular window within which the Compton photons are emitted.
In general, it follows from $\delta^{(1)}(P_N^-)$ that is present in Eq.~\eqref{cp5},
that $p_{\rm i}^0-p_{\rm i}^\parallel >K^0-K^\parallel$
which imposes the limits on the scattering angle of Compton photons. In particular, in the
electron rest frame this condition can be rewritten in a simple form,
\begin{equation}
 \cos\Theta_{\bm K}>1-m_{\rm e}c^2/\omega_{\bm K} .
\end{equation}
Hence, we find that for the parameters specified in Fig.~\ref{smooth2and32mod2a} ($\hbar\omega_{\bm K}=100 m_{\rm e}c^2$), 
the Compton photons are emitted in a cone such that $|\Theta_{\bm K}|< \sqrt{2}/10$
radians. This agrees well with our numerical results.

\section{Conclusions}
\label{conclusions}

In this paper, the nonlinear Compton scattering of an intense laser beam by an
electron has been considered. We have modeled the incident laser beam as a finite-length 
plane-wave-fronted laser pulse or a sequence of such pulses. For both cases, the frequency
and angular distributions of energy emitted as Compton photons have been unambiguously
defined such that a comparison between these two cases became meaningful. The theory of nonlinear
Compton scattering that has been developed here accounts for, in principle, arbitrary laser pulses 
(i.e., arbitrary pulse shapes, pulse durations, etc.) which are used nowadays in experiments.

Here, we have envisaged the case when the initial electron is at rest and it is stroked by
an intense laser beam. Comparing situations when the laser beam is represented by an 
individual laser pulse and by a train of such pulses we recognized that for 
pulses which include several oscillations of the laser
field, both approaches lead to seemingly same results for Compton radiation spectra.
This agreement deteriorates however with decreasing duration of the incident
pulse, and particularly for very-few-cycle pulses a detailed treatment of their 
spectral characteristic becomes important. A sensitivity of angular distributions
of emitted photons to the incoming pulse duration was also investigated.
We found that for ultrashort laser pulses the respective angular distributions
show asymmetries which are due to asymmetries of the vector potential defining
the incident laser beam. Those asymmetries disappear however for very long pulse
durations.

\section*{Acknowledgments}

This work is supported by the Polish National Science Center (NCN) under Grant No. 2011/01/B/ST2/00381.

\appendix

\section{Light-cone coordinates}
\label{LC}

Throughout the paper, we apply the following definitions and notations for the light-cone 
coordinates. If the space direction is determined by a normalized to 1 space vector $\bm{n}$, 
then for an arbitrary four-vector $a$ we introduce the light-cone coordinates such that
\begin{equation}
a^\|=\bm{n}\cdot\bm{a},\, a^-=a^0-a^\|, \, a^+=\frac{a^0+a^\|}{2}, \, \bm{a}^\bot=\bm{a}-a^\|\bm{n}.
\end{equation}
With this definitions,
\begin{equation}
a\cdot b=a^0b^0-\bm{a}\cdot\bm{b}=a^-b^+ + a^+b^- -\bm{a}^\bot\cdot\bm{b}^\bot ,
\end{equation}
and
\begin{equation}
\mathrm{d}^4x=\mathrm{d}x^-\mathrm{d}x^+\mathrm{d}^2\bm{x}^\bot.
\end{equation}

\section{Boca-Florescu transformation}
\label{BF}

Let $h(\phi)$ be a real and integrable function which does not vanish only within a finite interval, 
$\Phi_0\leqslant\phi\leqslant\Phi$. We define also the integral,
\begin{equation}
H(\phi)=\int_{-\infty}^\phi\mathrm{d}u\,h(u).
\end{equation}
Let us note that $H(\phi)$ is 0 when $\phi < \Phi_0$, it is a constant if $\phi>\Phi$ [we shall
further denote this constant as $H(\Phi)=\int_{\Phi_0}^\Phi \mathrm{d}u\,h(u)$], and rather
than that it can have any value.
We consider the integral,
\begin{equation}
C=\int\mathrm{d}^{4}{x}\,\mathrm{e}^{-\mathrm{i}H(k\cdot x)-\mathrm{i}Q\cdot x},
\end{equation}
in which $Q$ is an arbitrary four-vector and $k$ satisfies the zero-mass on-shell relation, $k\cdot k=0$. 
Let us assume that $k=k^0(1,\bm{n})$, where $\bm{n}^2=1$ and $k^0>0$. Hence, using the 
light-cone variables introduced in the Appendix \ref{LC}, we obtain that
\begin{equation}
C=(2\pi)^3\delta^{(1)}(Q^-)\delta^{(2)}(\bm{Q}^\bot)I(k^0,Q^+),
\end{equation}
where
\begin{equation}
I(k^0,Q^+)=\int\mathrm{d}{x^-}\mathrm{e}^{-\mathrm{i}[ H(k^0x^-)+Q^+x^-]}.
\end{equation}
Next, we regularize this integral similar to a regularization
introduced in Ref.~\cite{Boca1},
\begin{equation}
I_\varepsilon (k^0,Q^+)=\int\mathrm{d}{x^-} \mathrm{e}^{-\mathrm{i}[H(k^0x^-)+Q^+x^-]-\varepsilon |x^-|}, 
\end{equation}
so that it becomes absolutely convergent. Here, it is understood that $\varepsilon >0$. By noting that
\begin{align}
\int & \mathrm{d}{x^-}[-\mathrm{i}(k^0 h(k^0x^-)+Q^+)-\varepsilon \mathrm{sgn}(x^-)] \nonumber \\
 &\qquad\times\mathrm{e}^{-\mathrm{i}[H(k^0x^-)+Q^+x^-]-\varepsilon |x^-|} \nonumber \\ 
=&\mathrm{e}^{-\mathrm{i}[H(k^0x^-)+Q^+x^-]-\varepsilon |x^-|}\Big|_{-\infty}^{+\infty}=0, 
\end{align}
where $\mathrm{sgn}(x^-)$ is the sign function, we arrive at
\begin{equation}
\mathrm{i}Q_+ I_\varepsilon (k^0,Q_+)=I^{(1)}_\varepsilon+I^{(2)}_\varepsilon ,
\end{equation}
with
\begin{equation}
I^{(1)}_\varepsilon=-\mathrm{i}k^0\int\mathrm{d}{x^-} h(k^0x^-)\mathrm{e}^{-\mathrm{i}[H(k^0x^-)+Q^+x^-]-\varepsilon |x^-|},
\end{equation}
and
\begin{equation}
I^{(2)}_\varepsilon=-\varepsilon\int\mathrm{d}{x^-}\mathrm{sgn}(x^-)\mathrm{e}^{-\mathrm{i}[H(k_0x^-)+Q^+x^-]-\varepsilon |x^-|} .
\end{equation}
Since the integral $I^{(1)}_\varepsilon$ is over the finite region such that $\Phi^0/k^0 \leqslant x^-\leqslant \Phi/k^0$, 
we can put there $\varepsilon=0$. In the second integral we divide the integration space into three regions, 
$(-\infty,\Phi^0/k^0]$, $[\Phi^0/k^0,\Phi/k^0]$, and $[\Phi/k^0,\infty)$. This leads us to
\begin{align}
I^{(2)}_\varepsilon=&\theta(-\Phi_0)\mathrm{e}^{(\varepsilon-\mathrm{i}Q^+)\Phi^0/k^0} \frac{\varepsilon}{\varepsilon-\mathrm{i}Q^+} \\ 
+&\theta(\Phi_0)\biggl[\frac{\varepsilon}{\varepsilon-{\rm i}Q^+}+\frac{\varepsilon}{\varepsilon+{\rm i}Q^+}\Bigl(\mathrm{e}^{-(\varepsilon+\mathrm{i}Q^+)\Phi^0/k^0}-1\Bigr)\biggr]\nonumber\\
-&\varepsilon\int_{\Phi^0/k^0}^{\Phi/k^0}\mathrm{d}{x^-}\mathrm{sgn}(x^-) \mathrm{e}^{-\mathrm{i}[H(k^0x^-)+Q^+x^-]-\varepsilon |x^-|} \nonumber \\ 
+&\theta(-\Phi){\rm e}^{-{\rm i}H(\Phi)} \biggl[\frac{\varepsilon}{\varepsilon-{\rm i}Q^+}\Bigl(1-\mathrm{e}^{(\varepsilon-\mathrm{i}Q^+)\Phi^0/k^0}\Bigr)\nonumber\\
-&\frac{\varepsilon}{\varepsilon+{\rm i}Q^+}\biggr]-\theta(\Phi)\mathrm{e}^{-\mathrm{i}H(\Phi)-(\varepsilon+\mathrm{i}Q^+)\Phi/k^0} \frac{\varepsilon}{\varepsilon+\mathrm{i}Q^+},\nonumber
\end{align}
where $\theta(\Phi)$ is the step function. As one can see, 
this integral in the limit when $\varepsilon\rightarrow 0$ vanishes provided that $Q^+ \neq 0$. 
Since from the condition $\delta^{(1)}(Q^-)$ we have also $Q^0=Q^\|$, we conclude that $Q^+=Q^0 \neq 0$. Finally, we obtain
\begin{multline}
\int\mathrm{d}^{4}{x}\,\mathrm{e}^{-\mathrm{i}H(k\cdot x)-\mathrm{i}Q\cdot x} \\
=-\frac{k^0}{Q^0}\int\mathrm{d}^{4}{x} 
\,h(k\cdot x)\mathrm{e}^{-\mathrm{i}H(k\cdot x)-\mathrm{i}Q\cdot x} ,  \label{BocaFlorescu}
\end{multline}
which holds only if $Q^0\neq 0$. This is an analogue of the transformation derived by Boca and Florescu in Ref.~\cite{Boca1}.

\end{document}